# Modified Stage-Gate: A Conceptual Model of Virtual Product Development Process

Nader Ale Ebrahim[*], Shamsuddin Ahmed and Zahari Taha

Department of Engineering Design and Manufacture,

Faculty of Engineering, University of Malaya

50603 Lembah Pantai, Kuala Lumpur, Malaysia

*Correspondent author - email: aleebrahim@perdana.um.edu.my

*Abstract* – In today's dynamic marketplace, manufacturing companies are under strong pressure to introduce new products for long-term survival with their competitors. Nevertheless, every company cannot cope up progressively or immediately with the market requirements due to knowledge dynamics being experienced in competitive milieu. Increased competition and reduced product life cycles put force upon companies to develop new products faster. In response to these pressing needs there should be some new approach compatible in flexible circumstances. This paper presents a solution based on the popular Stage-Gate system, which is closely linked with virtual team approach. Virtual teams can provide a platform to advance the knowledge-base in a company and thus to reduce time-to-market. This article introduces conceptual product development architecture under a virtual team umbrella. The paper describes all the major aspects of new product development (NPD), NPD process and its relationship with virtual team, Stage-Gate system and finally presents a modified Stage-Gate system to cope up with the changing needs. It also provides the guidelines for the successful implementation of virtual team in new products development.

*Keywords* – Modified Stage-Gate System, Virtual Product Development, Conceptual Model

**1.0  INTRODUCTION**

New product development (NPD) is widely recognized as a key to corporate prosperity (Lam et al., 2007). Different products may require different processes, a new product idea needs to be conceived, selected, developed, tested and launched to the market (Martinez-Sanchez et al., 2006). The specialized skills and talents required for the development of new products often reside (and develop) locally in pockets of excellence around the company or even around the world. Firms therefore, have no choice but to disperse their new product units to access such dispersed knowledge and skills (Kratzer et al., 2005). As a result, firms are finding that internal development of all technology needed for new products and processes are difficult or impossible. They must increasingly acquire technology from external sources (Stock and Tatikonda, 2004).

Virtualization in NPD has recently started to make serious headway due to developments in technology - virtuality in NPD is now technically possible (Leenders et al., 2003). Automotive OEMs (Original equipment manufacturers) have formed partnerships with suppliers to take advantage of their technological expertise in development, design, and manufacturing (Wagner and Hoegl, 2006). As product development becomes more complex, supply chain also have to collaborate more closely than in the past. These kinds of collaborations almost always involve individuals from different locations, so virtual team working supported by IT, offers considerable potential benefits (Anderson et al., 2007). May and Carter (2001) in their case study of virtual team working in the European automotive industry have shown that enhanced communication and collaboration between geographically distributed engineers at automotive manufacturer and supplier sites make them get benefits in terms of better quality, reduced costs and a reduction in the time-to-market (between 20 to 50 percent) for a new product vehicle.

Although the uses of the internet in NPD have received considerable attention in the literature, very little is written about the collaborative tool and virtual team implementation in

NPD. On the other hand, Stage-Gate system which defines different steps of product development has some criticism and according to extent of information and communication technology (ICT) need to modify. In forthcoming section the major aspects of new product development (NPD), NPD process and its relationship with virtual team, Stage-Gate system and finally presents a modified Stage-Gate system will be described.

## 2.0 NPD CALLS FOR VIRTUALITY

Product development definition used by different researchers in slightly different ways but generally it is the process that covers product design, production system design and product introduction processes and start of production (Johansen, 2005). A multidisciplinary approach is needed to be successful in launching new products and managing daily operations (Flores, 2006). In the NPD context, teams developing new products in turbulent environments encounter quick depreciation of technology and market knowledge due to rapidly changing customer needs, wants, and desires (Akgun et al., 2007). Adoption of collaborative engineering tools and technology (e.g., Web-based development systems for virtual team coordination) was significantly correlated with NPD profitability (Ettlie and Elsenbach, 2007). ICT enhance the NPD process by shortening distances and saving on costs and time (Vilaseca-Requena et al., 2007).

Kafouros et al. (2008) found that internationalization enhances a firm's capacity to improve performance through innovation. Since efficiency, effectiveness and innovation management have different and contradictory natures, it is very difficult to achieve an efficient and innovative network cooperative NPD (Chen et al., 2008b). Supplier involvement in NPD can also help the buying firm to gain new competencies, share risks, move faster into new markets, and conserve resources (Wagner and Hoegl, 2006).

New product development (NPD) has long been recognised as one of the corporate core functions (Huang et al., 2004). During the past 25 years NPD has increasingly been recognized as a critical factor in ensuring the continued existence of firms (Biemans, 2003). The rate of market growth and technological changes have accelerated in the past years and this turbulent environment requires new methods and techniques to bring successful new products to the marketplace (González and Palacios, 2002). Particularly for companies with short product life cycles, it is important to quickly and safely develop new products and new product platforms that fulfil reasonable demands on quality, performance, and cost (Ottosson, 2004). The world market requires short product development times (Starbek and Grum, 2002), and therefore, in order to successfully and efficiently get all the experience needed in developing new products and services, more and more organizations are forced to move from traditional face-to-face teams to virtual teams or adopt a combination between the two types of teams (Precup et al., 2006).

Given the complexities involved in organizing face-to-face interactions among team members and the advancements in electronic communication technologies, firms are turning toward employing virtual NPD teams (Jacobsa et al., 2005, Badrinarayanan and Arnett, 2008, Schmidt et al., 2001). IT improve NPD flexibility (Durmusoglu and Calantone, 2006). New product development requires the collaboration of new product team members both within and outside the firm (Martinez-Sanchez et al., 2006, McDonough et al., 2001, Ozer, 2000) and NPD teams are necessary in almost all businesses (Leenders et al., 2003). In addition, the pressure of globalized competition forces companies to face increased pressures to build critical mass, reach new markets, and plug skill gaps. Therefore, NPD efforts are increasingly being pursued across multiple nations through all forms of organizational arrangements (Cummings and Teng, 2003). Given the resulting differences in time zones and physical distances in such efforts, virtual NPD projects are receiving increasing attention (McDonough et al., 2001). The use of virtual teams for new product development is rapidly growing and organizations can be dependent on it to sustain competitive advantage (Taifi, 2007).

## 2.1 New product Development Process

New business formation activities vary in complexity and formality from day-to-day entrepreneurial or customer prospecting activities to highly structured approaches to new product development (Davis and Sun, 2006). Today's uncertain and dynamic environment presents a fundamental challenge to the new product development process of the future (MacCormack et al., 2001). New product development is a multi-dimensional process and involves multiple activities (Ozer, 2000). Kusar al. (2004) summarized different stage of new product development which in earlier stages, the objective is to make a preliminary market, business, and technical assessment whereas at the later stages they propose to actually design and develop the product(s).

- Definition of goals (goals of the product development process)
- Feasibility study (term plan, financial plan, pre-calculation, goals of market)
- Development (first draft and structure of the product, first draft of components, product planning and its control processes)
- Design (design of components, drawing of parts, bills of material)

### 2.1.1 Stage-Gate System in NPD

Several authors proposed different conceptual models for the NPD process, beginning from the idea screening and ending with the commercial launching. The model of Cooper, called the Stage-Gate System is one of the most widely acknowledged systems (Rejeb et al., 2008). The Stage-Gate System model (Figure 1) divides the NPD into discrete stages, typically five stages. Each Stage gathers a set of activities to be done by a multifunctional project team. To enter into each stage, some conditions and criteria have to be fulfilled. These are specified in the Gates. A Gate is a project review in which all the information is confronted by the whole team. Some criticism of the method has surfaced, claiming that the steering group assessment in the stage and gate steps halts the project for an unnecessarily long time, making the process abrupt and discontinuous (Ottosson, 2004). A closer integration of management through virtual team in the process might be a solution for avoiding such situations.

### 2.1.2 Stage-Gate Process

This process is a method of managing the new product development process to increase the probability of launching new products quickly and successfully. The process provides a blueprint to move projects through the various stages of development: 1) idea generation, 2) preliminary investigation, 3) business case preparation, 4) product development, 5) product testing, and 6) product introduction. This process is used by such companies as IBM, Procter & Gamble, 3M, General Motors, and others. The process is primarily used in the development of specific commercial products, and is more likely to be used in platform projects than in derivative projects.

Auto companies that have modified their Stage-Gates procedures are also significantly more likely to report (1) use of virtual teams; (2) adoption of collaborative and virtual new product development software supporting tools; (3) having formalized strategies in place specifically to guide the new product development process; and (4) having adopted structured processes used to guide the new product development process (Ettlie and Elsenbach, 2007).

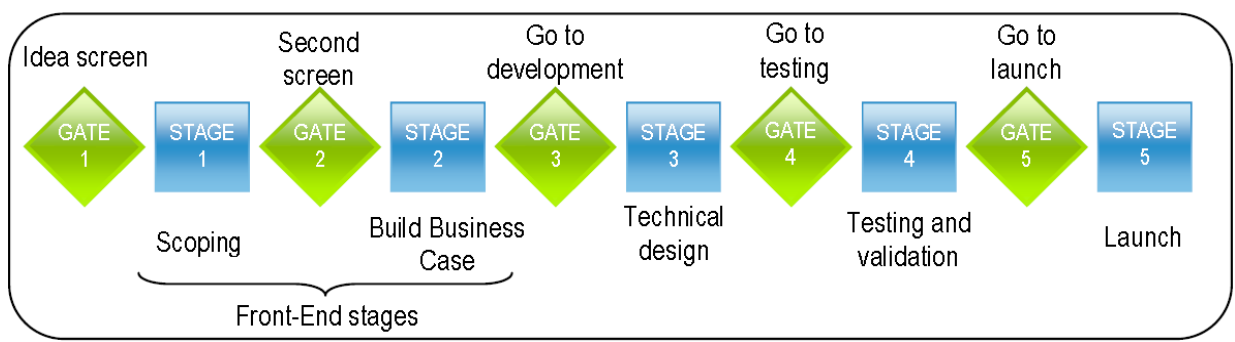

Figure 1 The Stage-Gate System model (source: Cooper, 2006)

## 3.0 DEMAND FOR MODIFIED STAGE-GATE WITH VIRTUAL PRODUCT DEVELOPMENT TEAM

Recently, the Stage-Gate system had been modified and adjusted to fit the real situation in nowadays, called the Next Generation Stage-Gate (Figure 2). The greatest change in Stage-Gate system is that it has become a scalable process, scaled to fit very different types and risk-levels of projects, from very risky and complex platform developments through to lower risk extensions and modifications, and even to handle rather simple sales force requests.

Managers recognized that any kinds of product development project have to manage risks and consumption of resources, but it is not all necessary to go through the fulfil five-stage process. The process has revised into multiple versions to fit business needs and to accelerate projects. Stage-Gate XPress for projects of moderate risk, such as improvements, modifications and extensions; and Stage-Gate Lite for very small projects, such as simple customer requests (Cooper, 2008). Although Next Generation Stage-Gate has defined for different types and risk-levels of projects, but still team collaboration in each stage is unveiled. So dealing with virtual team can bring an opportunity to make closer integration of team members in the process.

Virtual product development team by using collaborative tools can effectively be used both in the earlier and later stages of the NPD process. Past research has mainly focused on the role of Internet in NPD (Ozer, 2004). Almeida and Miguel (2007), have been identified in the literature that it seems to exist a lack of a conceptual model that represents all dimensions and interactions in the new product development process. On the other hand, some criticism of Stage-Gate method has surfaced, claiming that the steering group assessment in the gate step halts the project for an unnecessarily long time, making the process abrupt and discontinuous (Ottosson, 2004). A closer integration of management through virtual team in the process might be a solution for avoiding such situations. Integration is the essence of the concurrent product design and development activity in many organizations (Pawar and Sharifi, 1997). Ragatz et al.

(2002) suggest that integration of the supplier's technology roadmaps into the development cycle is critical to ensuring that target costs are met.

To compensate for the lack of conceptual model that represents all aspects and interactions in the new product process and decrease criticism of Stage-Gate system, a solution called Modified Stage-Gate system is introduced.

Figure 3 illustrates new model architecture of virtual product development process. The architecture is structured in a two-layered framework: Traditional Stage-Gate system and collaborative tool layer which is supported by virtual team. Merge of Stage-gate system with virtual product development team lead to increased new product performance and decreased time-to-market. The following sections will describe some elements of the collaborative tool layer in more detail.

Gassmann and Von Zedtwitz (2003) defined "virtual team as a group of people and sub-teams who interact through interdependent tasks guided by common purpose and work across links strengthened by information, communication, and transport technologies." Another definition suggests that virtual teams, are distributed work teams whose members are geographically dispersed and coordinate their work predominantly with electronic information and communication technologies (e-mail, video-conferencing, telephone, etc.) (Hertel et al., 2005). We define, virtual team is small temporary groups of geographically, organizationally and/or time dispersed knowledge workers who coordinate their work predominantly with electronic information and communication technologies in order to accomplish one or more organization tasks.

## 3.1 Capturing Customer Requirements

Collaborative tools allow firms to respond quickly to specific customer requirements with new, high-quality, innovative products, and it enables firms to build cross-functional competencies, enhance flexibility and share knowledge (Mulebeke and Zheng, 2006). Capturing customer requirements is represented throughout product development will facilitate performing quality function deployment (Rodriguez and Al-Ashaab, 2005).

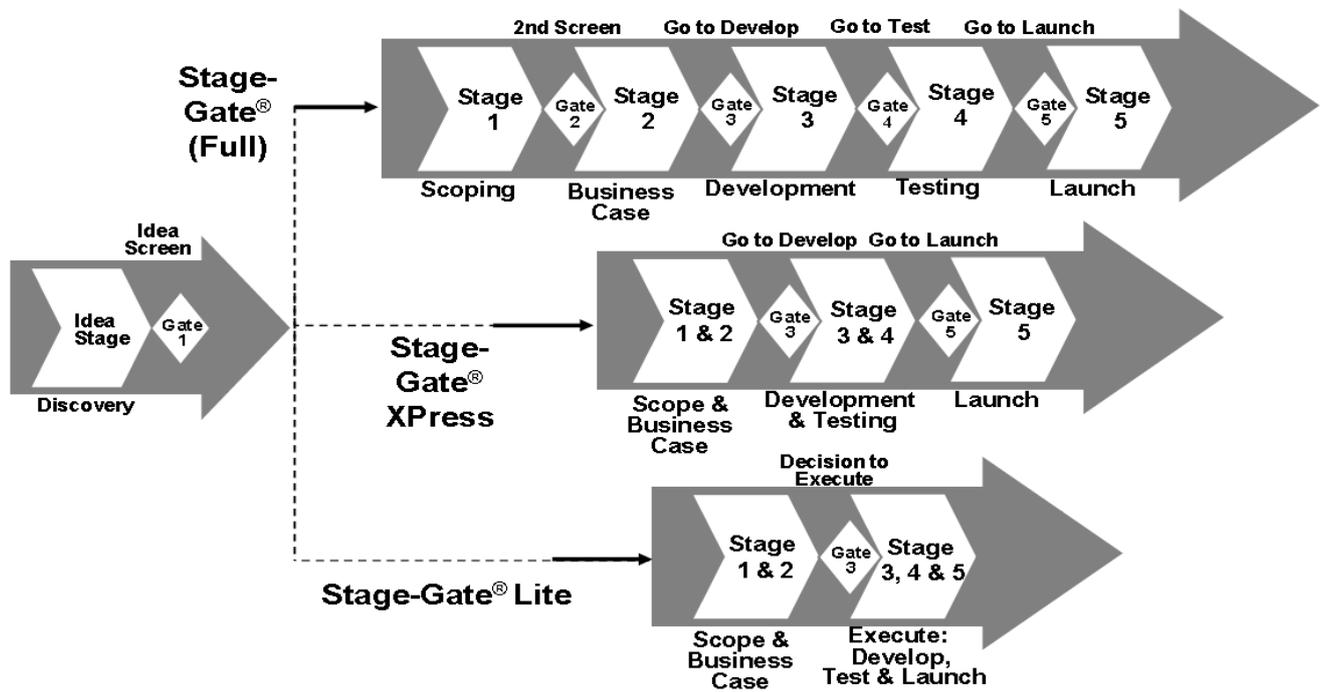

Figure 2  An overview of Next Generation Stag-Gate (Source: (Cooper, 2008))

### 3.2  Collaborative Capabilities

Enabling collaborative capability through virtual teamwork represents a fundamental transitioning to more effective organizational work practices (Susman et al., 2003).

The use of virtual team will change the communication pattern both within and outside the firm. Successful collaborations require more than the mere use of electronic communication and involve new skills and a supportive context that provides commitment and resources to facilitate collaboration (Martinez-Sanchez et al., 2006).

### 3.3  Company Resources

Virtual team provides cost savings to employees by eliminating time-consuming commutes to central offices and offers employees more flexibility to co-ordinate their work and family responsibilities (Johnson et al., 2001). Virtual teams overcome the limitations of time, space, and organizational affiliation that traditional teams face (Piccoli et al., 2004) and able to digitally or electronically unite experts in highly specialized fields working at great distances from each other (Rosen et al., 2007).

Top management support is a strong motivational factor in the entire new product process. Although collaborative tools are able to assists top management but many managers are uncomfortable with the concept of a virtual team because successful management of virtual teams may require new methods of supervision (Jarvenpaa and Leidner, 1999). Management commitment provides organizational support for change, generates enthusiasm, provides a clear vision of the product concept and assures sufficient allocation of resources (González and Palacios, 2002).

Information sharing has been identified as an important success factor in NPD (Ozer, 2006). The positive impact of information sharing on the success of new products has long been established in the NPD literature (Sridhar et al., 2007, Furst et al., 2004, Merali and Davies, 2001, Lipnack and Stamps, 2000).

Virtual teams reduce time-to-market (Sorli et al., 2006, Kankanhalli et al., 2006, Chen, 2008, Shachaf, 2008, Ge and Hu, 2008, Guniš et al., 2007). Lead time or time to market has been generally admitted to be one of the most important keys for success in manufacturing companies (Sorli et al., 2006). Time also has an almost 1:1 correlation with cost, so cost will likewise be reduced if the time-to market is quicker (Rabelo and Jr., 2005). Virtual teams overcome the limitations of time, space, and organizational affiliation that traditional teams face (Piccoli et al., 2004) and reducing relocation time and costs, reduced travel costs (Bergiel et al., 2008, Fuller et al., 2006, Kankanhalli et al., 2006, Olson-Buchanan et al., 2007). Virtual NPD teams overcome the limitations of time, space, and organizational affiliation that traditional teams face (Piccoli et al., 2004). Virtual R&D team is able to tap selectively into centre of excellence, using the best talent regardless of location (Criscuolo, 2005, Samarah et al., 2007, Fuller et al., 2006, Badrinarayanan and Arnett, 2008, Furst et al., 2004).

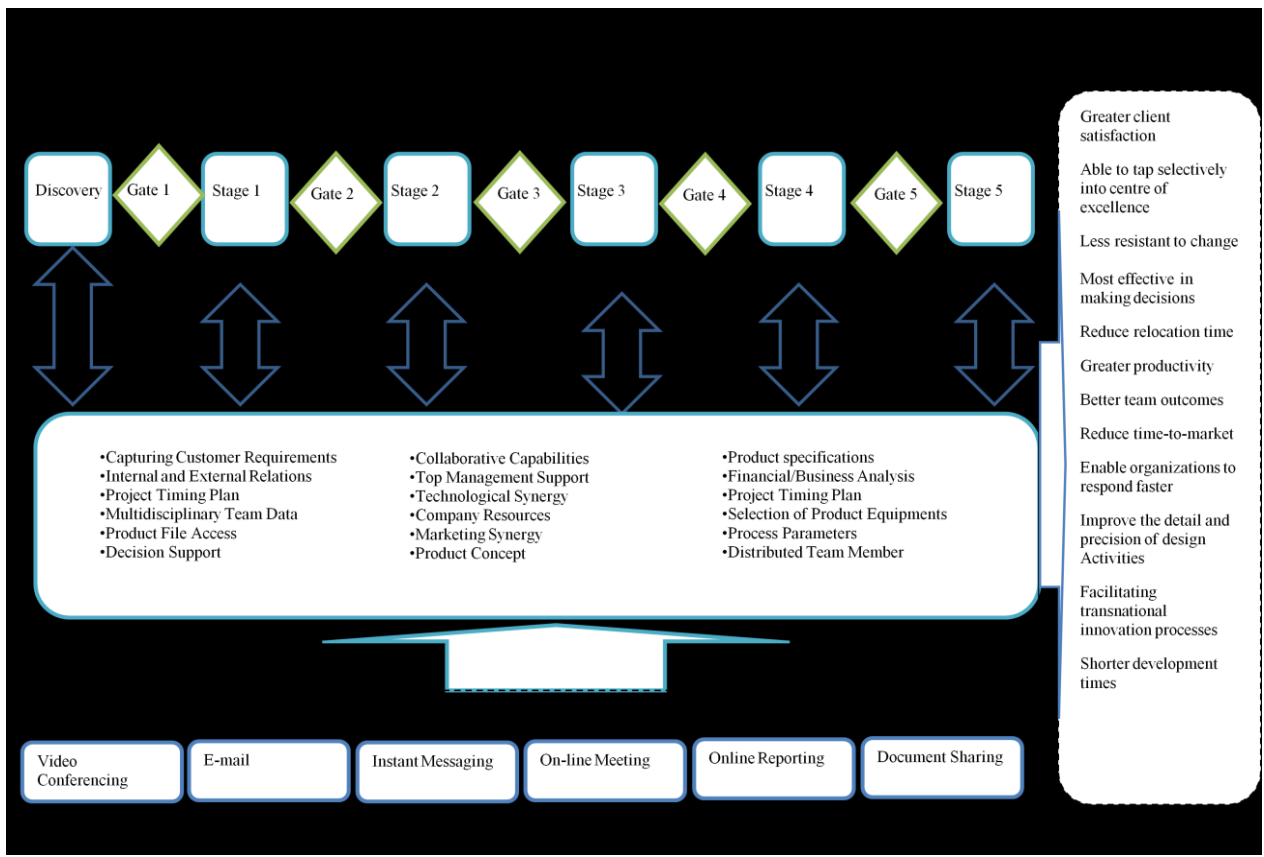

Figure 3 Modified Stage-Gate: Model architecture of Virtual product development Process

Virtual team also, respond quickly to changing business environments (Bergiel et al., 2008, Mulebeke and Zheng, 2006), able to digitally or electronically unite experts in highly specialized fields working at great distances from each other (Rosen et al., 2007), more effective R&D continuation decisions (Cummings and Teng, 2003, Schmidt et al., 2001), most effective in making decisions (Hossain and Wigand, 2004, Paul et al., 2004), provide greater degree of freedom to individuals involved with the development project (Ojasalo, 2008, Badrinarayanan and Arnett, 2008, Prasad and Akhilesh, 2002), Greater productivity, shorter development times (McDonough et al., 2001, Mulebeke and Zheng, 2006), Producing better outcomes and attract better employees, Generate the greatest competitive advantage from limited resources (Martins et al., 2004, Chen et al., 2008c, Rice et al., 2007), Useful for projects that require cross-functional or cross boundary skilled inputs (Lee-Kelley and Sankey, 2008), Less resistant to change (Precup et al., 2006), Facilitating transnational innovation processes (Gassmann and Von Zedtwitz, 2003, Prasad and Akhilesh, 2002), higher degree of cohesion (Teams can be organized whether or not members are in proximity to one another) (Kratzer et al., 2005, Cascio, 2000, Gaudes et al., 2007), Evolving organizations from production-oriented to service/information-oriented (Johnson et al., 2001, Precup et al., 2006) and Provide organizations with unprecedented level of flexibility and responsiveness (Hunsaker and Hunsaker, 2008, Chen, 2008, Pihkala et al., 1999, Liu and Liu, 2007). Beside these advantages virtual NPD teams are self-assessed performance and high performance (Chudoba et al., 2005, Poehler and Schumacher, 2007), employees perform their work without concern of space or time constraints (Lurey and Raisinghani, 2001), optimize the contributions of individual members toward the completion of business tasks and organizational goal (Samarah et al., 2007), reduce the pollution (Johnson et al., 2001), manage the development and commercialization tasks quite well (Chesbrough and Teece, 2002), Improve communication and coordination, and encourage the mutual sharing of inter-organizational resources and competencies (Chen et al., 2008a), employees can more easily accommodate both personal and professional lives (Cascio, 2000), cultivating and managing creativity (Leenders et al., 2003, Atuahene-Gima, 2003, Badrinarayanan and Arnett, 2008), facilitate knowledge capture and sharing knowledge, experiences (Rosen et al., 2007, Zakaria et al., 2004, Furst et al., 2004, Sridhar et al., 2007), Improve the detail and precision of design activities (Vaccaro et al., 2008), Provide a vehicle for global collaboration and coordination of R&D-related activities (Paul et al., 2005 ), Allow organizations to access the most qualified individuals for a particular job regardless of their location (Hunsaker and Hunsaker, 2008) and Enable organizations to respond faster to increased competition (Hunsaker and Hunsaker, 2008, Pauleen, 2003).

The ratio of virtual R&D member publications exceeded from co-located publications (Ahuja et al., 2003) and The extent of informal exchange of information is minimal (Pawar and Sharifi, 1997, Schmidt et al., 2001). Virtual teams have better team outcomes (quality, productivity, and satisfaction) (Gaudes et al., 2007, Ortiz de Guinea et al., 2005, Piccoli et al., 2004), Reduce training expenses, Faster Learning (Pena-Mora et al., 2000, Atuahene-Gima, 2003, Badrinarayanan and Arnett, 2008) and finally greater client satisfaction (Jain and Sobek, 2006).

## 4.0 KEY FACTORS FOR SUCCESSFULLY IMPLEMENTING VIRTUAL TEAM IN NPD

NPD is continuing to be an area that is receiving increased attention, both in practice and academic spheres (Shani et al., 2003). Eppinger and Chitkara (2006) studied global product development (GPD) base on virtual team, for companies in the manufacturing sector by conducting interviews with 30 executives and surveying over 1150 product development executives and professionals from large manufacturing companies. They reported the following ten key success factors for successful GPD:

- Management priority and commitment – Commitment from management to make the necessary organization, process and cultural changes to make GPD work.
- Process modularity for global distribution – Ability to separate activities into modular work packages for global distribution.
- Product modularity to develop subsystems or components in different locations – Ability to break down into subsystems for global distribution.
- Core competence so the company does not become completely reliant on suppliers or contractors – Good understanding of what the company's core competencies are, so that do not get outsourced.
- Intellectual property, which becomes more difficult to protect – Defining process and products in a modular way to protect IP.
- Data quality, which concerns availability, accessibility, and audit ability – Ability to update and share data with teams in multiple locations.

- Infrastructure (including networks and power supplies) to support activities in all locations – Unified infrastructure, systems, technologies, and processes that are shared between all locations.
- Governance and product management is needed to coordinate and monitor the entire effort – Ability to coordinate and monitor program, including detailed project planning.
- Collaborative culture is necessary and is helped by a consistent set of processes and standards – Building and sustaining trust, ensuring teams have consistent processes and standards.

Organization change management requires planning, training, and education of those in key roles for global product development plan and train for new roles, behaviours, and skills.

## 5.0 CONCLUSION

The internet, incorporating computers and multimedia, have provided tremendous potential for remote integration and collaboration in business and manufacturing applications. Most companies today are divided in different departments located in different geographical places and dealing with specialized tasks. So using collaborative tools enables authorized users in geographically dispersed locations to have access to the company's product data and carry out product development work simultaneously and collaboratively on any operating systems.

The modified Stage-Gate system has demonstrated to be a good development platform for the NPD. In order to integrate and share the information and knowledge available within geographically distributed companies, this model can be a reference model. The proposed model architecture of virtual product development process, does not aim to replace the existing systems in companies but rather to be a support tool for communicating and sharing knowledge among the disperse partners. Modified Stage-Gate system will lead to the production of better and more cost effective products, developed in a shorter period of time.

In highly competitive era which forces companies to launch new product faster, the decision on setting up virtual teams and using a modified NPD process is not a choice but a requirement. The theme of virtual teams and application of collaborative tool in NPD has not been much explored and researchers in this field are encouraging more studies and analyses to be made.

**REFERENCES**

AHUJA, M. K., GALLETTA, D. F. & CARLEY, K. M. (2003) Individual Centrality and Performance in Virtual R&D Groups: An Empirical Study *Management Science,* 49**,** 21-38.

AKGUN, A. E., BYRNE, J. C., LYNN, G. S. & KESKIN, H. (2007) New product development in turbulent environments: Impact of improvisation and unlearning on new product performance. *Journal of Engineering and Technology Management,* 24**,** 203–230.

ALMEIDA, L. & MIGUEL, P. (2007) Managing new product development process: a proposal of a theoretical model about their dimensions and the dynamics of the process. *Complex Systems Concurrent Engineering.* Springer London.

ANDERSON, A. H., MCEWAN, R., BAL, J. & CARLETTA, J. (2007) Virtual team meetings: An analysis of communication and context. *Computers in Human Behavior,* 23**,** 2558–2580.

ATUAHENE-GIMA, K. (2003) The effects of centrifugal and centripetal forces on product development speed and quality: how does problem solving matter? . *Academy of Management Journal,* 46**,** 359-373.

BADRINARAYANAN, V. & ARNETT, D. B. ( 2008) Effective virtual new product development teams: an integrated framework. *Journal of Business & Industrial Marketing,* 23**,** 242-248.

BERGIEL, J. B., BERGIEL, E. B. & BALSMEIER, P. W. (2008) Nature of virtual teams: a summary of their advantages and disadvantages. *Management Research News,* 31**,** 99-110.

BIEMANS, W. G. (2003) A picture paints a thousand numbers: a critical look at b2b product development research *Business & Industrial Marketing,* 18**,** 514-528.

CASCIO, W. F. (2000) Managing a virtual workplace. *The Academy of Management Executive,*

14**,** 81-90.

CHEN, H. H., KANG, Y. K., XING, X., LEE, A. H. I. & TONG, Y. (2008a) Developing new products with knowledge management methods and process development management in a network. *Computers in Industry,* 59**,** 242–253.

CHEN, H. H., LEE, A. H. I., WANG, H. Z. & TONG, Y. (2008b) Operating NPD innovatively with different technologies under a variant social environment. *Technological Forecasting & Social Change***,** 385–404.

CHEN, T.-Y. (2008) Knowledge sharing in virtual enterprises via an ontology-based access control approach. *Computers in Industry,* Article In press**,** No of Pages 18.

CHEN, T. Y., CHEN, Y. M. & CH, H. C. (2008c) Developing a trust evaluation method between co-workers in virtual project team for enabling resource sharing and collaboration. *Computers in Industry* 59**,** 565-579.

CHESBROUGH, H. W. & TEECE, D. J. (2002) Organizing for Innovation: When Is Virtual Virtuous? *Harvard Business Review Article,* August 127-135.

CHUDOBA, K. M., WYNN, E., LU, M., WATSON-MANHEIM & BETH, M. (2005) How virtual are we? Measuring virtuality and understanding its impact in a global organization. *Information Systems Journal,* 15**,** 279-306.

COOPER, R. G. (2006) Managing Technology Development Projects. *Research Technology Management,* 49**,** 23-31.

COOPER, R. G. (2008) Perspective: The Stage-Gate® Idea-to-Launch Process—Update, What's New, and NexGen Systems. *Journal of Product Innovation Management,* 25**,** 213-232.

CRISCUOLO, P. (2005) On the road again: Researcher mobility inside the R&D network. *Research Policy,* 34**,** 1350–1365

CUMMINGS, J. L. & TENG, B. S. (2003) Transferring R&D knowledge: the key factors affecting

knowledge transfer success. *Journal of Engineering Technology Management*, 39–68.

DAVIS, C. H. & SUN, E. (2006) Business Development Capabilities in Information Technology SMEs in a Regional Economy: An Exploratory Study. *The Journal of Technology Transfer,* 31, 145-161.

DURMUSOGLU, S. S. & CALANTONE, R. J. (2006) Is more information technology better for new product development? *Product & Brand Management,* 15, 435-441.

EPPINGER, S. D. & CHITKARA, A. R. (2006) The New Practice of Global Product Development. *MIT Sloan Management Review,* 47, 22-30.

ETTLIE, J. E. & ELSENBACH, J. M. (2007) Modified Stage-Gate Regimes in New Product Development. *Journal of Product Innovation Management,* 24, 20-33.

FLORES, M. (2006) IFIP International Federation for Information Processing. *Network-Centric Collaboration and Supporting Fireworks.* Boston, Springer.

FULLER, M. A., HARDIN, A. M. & DAVISON, R. M. (2006) Efficacy in Technology-Mediated Distributed Team *Journal of Management Information Systems,* 23, 209-235.

FURST, S. A., REEVES, M., ROSEN, B. & BLACKBURN, R. S. (2004) Managing the life cycle of virtual teams. *Academy of Management Executive,* 18, 6-20.

GASSMANN, O. & VON ZEDTWITZ, M. (2003) Trends and determinants of managing virtual R&D teams. *R&D Management* 33, 243-262.

GAUDES, A., HAMILTON-BOGART, B., MARSH, S. & ROBINSON, H. (2007) A Framework for Constructing Effective Virtual Teams *The Journal of E-working* 1, 83-97

GE, Z. & HU, Q. (2008) Collaboration in R&D activities: Firm-specific decisions. *European Journal of Operational Research* 185, 864-883.

GONZáLEZ, F. J. M. & PALACIOS, T. M. B. (2002) The effect of new product development techniques on new product success in Spanish firms. *Industrial Marketing Management* 31, 261-271.

GUNIŠ, A., ŠIŠLáK, J. & VALČUHA, Š. (2007) Implementation Of Collaboration Model Within SME's. IN CUNHA, P. F. & MAROPOULOS, P. G. (Eds.) *Digital Enterprise Technology-Perspectives and Future Challenges.* Springer US.

HERTEL, G. T., GEISTER, S. & KONRADT, U. (2005) Managing virtual teams: A review of current empirical research. *Human Resource Management Review,* 15**,** 69–95.

HOSSAIN, L. & WIGAND, R. T. (2004) ICT Enabled Virtual Collaboration through Trust. *Journal of Computer-Mediated Communication,* 10.

HUANG, X., SOUTAR, G. N. & BROWN, A. (2004) Measuring new product success: an empirical investigation of Australian SMEs. *Industrial Marketing Management,* 33**,** 117–123.

HUNSAKER, P. L. & HUNSAKER, J. S. (2008) Virtual teams: a leader's guide. *Team Performance Management,* 14**,** 86-101.

JACOBSA, J., MOLL, J. V., KRAUSE, P., KUSTERS, R., TRIENEKENS, J. & BROMBACHER, A. (2005) Exploring defect causes in products developed by virtual teams *Information and Software Technology,* 47**,** 399-410.

JAIN, V. K. & SOBEK, D. K. (2006) Linking design process to customer satisfaction through virtual design of experiments. *Research in Engineering Design,* 17 59-71.

JARVENPAA, S. L. & LEIDNER, D. E. (1999) Communication and Trust in Global Virtual Teams. *Organization Science* 10**,** 791 - 815

JOHANSEN, K. (2005) Collaborative Product Introduction within Extended Enterprises. *Department of Mechanical Engineering.* Linköping, Sweden, Linköpings Universitet.

JOHNSON, P., HEIMANN, V. & O'NEILL, K. (2001) The "wonderland" of virtual teams. *Journal of Workplace Learning,* 13**,** 24 - 30.

KAFOUROS, M. I., BUCKLEY, P. J., SHARP, J. A. & WANG, C. (2008) The role of internationalization in explaining innovation performance. *Technovation,* 28**,** 63–74.

KANKANHALLI, A., TAN, B. C. Y. & WEI, K.-K. (2006) Conflict and Performance in Global Virtual Teams. *Journal of Management Information Systems,* 23**,** 237-274.

KRATZER, J., LEENDERS, R. & ENGELEN, J. V. (2005) Keeping Virtual R&D Teams Creative. *Industrial Research Institute, Inc.,* March-April**,** 13-16.

KUSAR, J., DUHOVNIK, J., GRUM, J. & STARBEK, M. (2004) How to reduce new product development time. *Robotics and Computer-Integrated Manufacturing* 20**,** 1-15.

LAM, P.-K., CHIN, K.-S., YANG, J.-B. & LIANG, W. (2007) Self-assessment of conflict management in client-supplier collaborative new product development. *Industrial Management & Data Systems,* 107**,** 688 - 714.

LEE-KELLEY, L. & SANKEY, T. (2008) Global virtual teams for value creation and project success: A case study. *International Journal of Project Management* 26**,** 51–62.

LEENDERS, R. T. A. J., ENGELEN, J. M. L. V. & KRATZER, J. (2003) Virtuality, communication, and new product team creativity: a social network perspective. *Journal of Engineering and Technology Management,* 20**,** 69–92.

LIPNACK, J. & STAMPS, J. (2000) Why The Way to Work. *Virtual Teams: People Working across Boundaries with Technology.* Second Edition ed. New York, John Wiley & Sons.

LIU, B. & LIU, S. (2007) Value Chain Coordination with Contracts for Virtual R&D Alliance Towards Service. *The 3rd IEEE International Conference on Wireless Communications, Networking and Mobile Computing, WiCom 2007.* Shanghai, China, IEEE Xplore.

LUREY, J. S. & RAISINGHANI, M. S. (2001) An empirical study of best practices in virtual teams *Information & Management,* 38**,** 523-544.

MACCORMACK, A., VERGANTI, R. & IANSITI, M. (2001) Developing Products on "Internet Time": The Anatomy of a Flexible Development Process. *MANAGEMENT SCIENCE,* 47**,** 133-150.

MARTINEZ-SANCHEZ, A., PEREZ-PEREZ, M., DE-LUIS-CARNICER, P. & VELA-JIMENEZ, M.

J. (2006) Teleworking and new product development. *European Journal of Innovation Management,* 9**,** 202-214.

MARTINS, L. L., GILSON, L. L. & MAYNARD, M. T. (2004) Virtual teams: What do we know and where do we go from here? *Journal of Management,* 30**,** 805–835.

MAY, A. & CARTER, C. (2001) A case study of virtual team working in the European automotive industry. *International Journal of Industrial Ergonomics,* 27**,** 171-186.

MCDONOUGH, E. F., KAHN, K. B. & BARCZAK, G. (2001) An investigation of the use of global, virtual, and collocated new product development teams. *The Journal of Product Innovation Management,* 18**,** 110–120.

MERALI, Y. & DAVIES, J. (2001) Knowledge Capture and Utilization in Virtual Communities. *International Conference On Knowledge Capture, K-CAP'01.* Victoria, British Columbia, Canada.

MULEBEKE, J. A. W. & ZHENG, L. (2006) Incorporating integrated product development with technology road mapping for dynamism and innovation. *International Journal of Product Development*  3**,** 56 - 76.

OJASALO, J. (2008) Management of innovation networks: a case study of different approaches. *European Journal of Innovation Management,* 11**,** 51-86.

OLSON-BUCHANAN, J. B., RECHNER, P. L., SANCHEZ, R. J. & SCHMIDTKE, J. M. (2007) Utilizing virtual teams in a management principles course. *Education + Training,* 49**,** 408-423.

ORTIZ DE GUINEA, A., WEBSTER, J. & STAPLES, S. ( 2005) A Meta-Analysis of the Virtual Teams Literature. *Symposium on High Performance Professional Teams Industrial Relations Centre.* School of Policy Studies, Queen's University, Kingston, Canada.

OTTOSSON, S. (2004) Dynamic product development -- DPD. *Technovation,* 24**,** 207-217.

OZER, M. (2000) Information Technology and New Product Development Opportunities and

Pitfalls. *Industrial Marketing Management* 29**,** 387-396.

OZER, M. (2004) The role of the Internet in new product performance: A conceptual investigation. *Industrial Marketing Management* 33**,** 355– 369.

OZER, M. (2006) New product development in Asia: An introduction to the special issue. *Industrial Marketing Management,* 35**,** 252-261.

PAUL, S., SEETHARAMAN, P., SAMARAH, I. & MYKYTYN, P. P. (2004) Impact of heterogeneity and collaborative conflict management style on the performance of ynchronous global virtual teams. *Information & Management,* 41**,** 303-321.

PAUL, S., SEETHARAMAN, P., SAMARAH, I. & PETER MYKYTYN, J. ( 2005 ) Understanding Conflict in Virtual Teams: An Experimental Investigation using Content Analysis. *38th Hawaii International Conference on System Sciences.* Hawaii.

PAULEEN, D. J. (2003) An Inductively Derived Model of Leader-Initiated Relationship Building with Virtual Team Members. *Journal of Management Information Systems,* 20**,** 227-256.

PAWAR, K. S. & SHARIFI, S. (1997) Physical or virtual team collocation: Does it matter? *International Journal of Production Economics* 52**,** 283-290.

PENA-MORA, F., HUSSEIN, K., VADHAVKAR, S. & BENJAMIN, K. (2000) CAIRO: a concurrent engineering meeting environment for virtual design teams. *Artifcial Intelligence in Engineering* 14**,** 203-219.

PICCOLI, G., POWELL, A. & IVES, B. (2004) Virtual teams: team control structure, work processes, and team effectiveness. *Information Technology & People,* 17**,** 359 - 379.

PIHKALA, T., VARAMAKI, E. & VESALAINEN, J. (1999) Virtual organization and the SMEs: a review and model development. *Entrepreneurship & Regional Development,* 11**,** 335 - 349.

POEHLER, L. & SCHUMACHER, T. (2007) The Virtual Team Challenge: Is It Time for Training? *PICMET 2007.* Portland, Oregon - USA

PRASAD, K. & AKHILESH, K. B. (2002) Global virtual teams: what impacts their design and performance? *Team Performance Management,* 8**,** 102 - 112.

PRECUP, L., O'SULLIVAN, D., CORMICAN, K. & DOOLEY, L. (2006) Virtual team environment for collaborative research projects. *International Journal of Innovation and Learning,* 3**,** 77 - 94

RABELO, L. & JR., T. H. S. (2005) Sustaining growth in the modern enterprise: A case study. *Jornal of Engineering and Technology Management JET-M,* 22 274-290.

RAGATZ, G. L., HANDFIELD, R. B. & PETERSEN, K. J. (2002) Benefits associated with supplier integration into new product development under conditions of technology uncertainty. *Journal of Business Research,* 55**,** 389-400.

REJEB, H. B., MOREL-GUIMARAES, L. & BOLY, V. (2008) A new methodology based on Kano Model for needs evaluation and innovative concepts comparison during the front-end phases. *The Third European Conference on Management of Technology, EUROMOT 2008.* Nice, France.

RICE, D. J., DAVIDSON, B. D., DANNENHOFFER, J. F. & GAY, G. K. (2007) Improving the Effectiveness of Virtual Teams by Adapting Team Processes. *Computer Supported Cooperative Work (CSCW),* 16**,** 567-594.

RODRIGUEZ, K. & AL-ASHAAB, A. (2005) Knowledge web-based system architecture for collaborative product development. *Computers in Industry,* 56**,** 125-140.

ROSEN, B., FURST, S. & BLACKBURN, R. (2007) Overcoming Barriers to Knowledge Sharing in Virtual Teams. *Organizational Dynamics,* 36**,** 259–273.

SAMARAH, I., PAUL, S. & TADISINA, S. (2007) Collaboration Technology Support for Knowledge Conversion in Virtual Teams: A Theoretical Perspective. *40th Hawaii International Conference on System Sciences (HICSS).* Hawai.

SCHMIDT, J. B., MONTOYA-WEISS, M. M. & MASSEY, A. P. (2001) New product development

decision-making effectiveness: Comparing individuals, face-to-face teams, and virtual teams. *Decision Sciences,* 32**,** 1-26.

SHACHAF, P. (2008) Cultural diversity and information and communication technology impacts on global virtual teams: An exploratory study. *Information & Management,* 45**,** 131-142.

SHANI, A. B., SENA, J. A. & OLIN, T. (2003) Knowledge management and new product development: a study of two companies. *European Journal of Innovation Management,* 6**,** 137-149.

SORLI, M., STOKIC, D., GOROSTIZA, A. & CAMPOS, A. (2006) Managing product/process knowledge in the concurrent/simultaneous enterprise environment. *Robotics and Computer-Integrated Manufacturing,* 22**,** 399–408.

SRIDHAR, V., NATH, D., PAUL, R. & KAPUR, K. (2007) Analyzing Factors that Affect Performance of Global Virtual Teams. *Second International Conference on Management of Globally Distributed Work* Indian Institute of Management Bangalore, India.

STARBEK, M. & GRUM, J. (2002) Concurrent engineering in small companies. *International Journal of Machine Tools and Manufacture,* 42**,** 417-426.

STOCK, G. N. & TATIKONDA, M. V. (2004) External technology integration in product and process development. *International Journal of Operations & Production Management,* 24**,** 642-665.

SUSMAN, G. I., GRAY, B. L., PERRY, J. & BLAIR, C. E. (2003) Recognition and reconciliation of differences in interpretation of misalignments when collaborative technologies are introduced into new product development teams. *Journal of Engineering and Technology Management,* 20**,** 141–159.

TAIFI, N. (2007) Organizational Collaborative Model of Small and Medium Enterprises in the Extended Enterprise Era: Lessons to Learn from a Large Automotive Company and its dealers' Network. *Proceedings of the 2nd PROLEARN Doctoral Consortium on*

*Technology Enhanced Learning, in the 2nd European Conference on Technology Enhanced Learning.* Crete, Greece, CEUR Workshop Proceedings.

VACCARO, A., VELOSO, F. & BRUSONI, S. (2008) The Impact of Virtual Technologies on Organizational Knowledge Creation: An Empirical Study. *Hawaii International Conference on System Sciences.* Proceedings of the 41st Annual Publication

VILASECA-REQUENA, J., TORRENT-SELLENS, J. & JIMÉNEZ-ZARCO, A. I. (2007) ICT use in marketing as innovation success factor-Enhancing cooperation in new product development processes. *European Journal of Innovation Management,* 10**,** 268-288.

WAGNER, S. M. & HOEGL, M. (2006) Involving suppliers in product development: Insights from R&D directors and project managers. *Industrial Marketing Management,* 35**,** 936–943.

ZAKARIA, N., AMELINCKX, A. & WILEMON, D. (2004) Working Together Apart? Building a Knowledge-Sharing Culture for Global Virtual Teams. *Creativity and Innovation Management,* 13**,** 15-29.